\documentclass[preprintnumbers,amsmath,amssymbm,prd]{revtex4}
\usepackage{epsfig}
\usepackage{graphicx}

\begin{document}
\title{Strong cosmic censorship in charged black-hole spacetimes: As strong as ever}
\author{Shahar Hod}
\affiliation{The Ruppin Academic Center, Emeq Hefer 40250, Israel}
\affiliation{ }
\affiliation{The Hadassah Institute, Jerusalem 91010, Israel}
\date{\today}

\begin{abstract}
\ \ \ It is proved that dynamically formed Reissner-Nordstr\"om-de
Sitter (RNdS) black holes, which have recently been claimed to
provide counter-examples to the Penrose strong cosmic censorship
conjecture, are characterized by unstable (singular) inner Cauchy
horizons. The proof is based on analytical techniques which
explicitly reveal the fact that {\it charged} massive scalar fields
in the charged RNdS black-hole spacetime are characterized, in the
large-coupling regime, by quasinormal resonant frequencies with
$\Im\omega^{\text{min}}<{1\over2}\kappa_+$, where $\kappa_+$ is the
surface gravity of the black-hole event horizon. This result implies
that the corresponding relaxation rate $\psi\sim
e^{-\Im\omega^{\text{min}}t}$ of the collapsed charged fields is
slow enough to guarantee, through the mass-inflation mechanism, the
instability of the dynamically formed inner Cauchy horizons. Our
results reveal the physically important fact that, taking into
account the unavoidable presence of {\it charged} matter fields in
dynamically formed {\it charged} spacetimes, non-asymptotically flat
RNdS black holes are globally hyperbolic and therefore respect the
fundamental strong cosmic censorship conjecture.
\end{abstract}
\bigskip
\maketitle


{\it Introduction.---} The physically important and mathematically
elegant singularity theorems of Hawking and Penrose
\cite{HawPen,Pen}, published almost five decades ago, have forever
changed our understanding of the physical properties of curved
spacetimes. In particular, these theorems have revealed the
intriguing fact that spacetimes singularities, regions in which
classical theories of gravity lose their predictive power, may
naturally be formed from the dynamical gravitational collapse of
compact enough matter configurations.

In order to maintain the utility of general relativity in
successfully describing gravitational phenomena in our universe,
Penrose \cite{Pen} has put forward a physically intriguing
conjecture, known as the cosmic censorship hypothesis. In its strong
version, the conjecture asserts that, starting with physically
reasonable generic initial data, the dynamics of self-gravitating
physical systems, which are governed by the non-linearly coupled
Einstein-matter field equations, will always produce globally
hyperbolic spacetimes. The strong cosmic censorship (SCC) conjecture
therefore implies that, in curved spacetimes that can be formed
dynamically from physically acceptable initial conditions,
singularities only appear on spacelike or null hypersurfaces and
thus, until the very moment of encounter, a spacetime singularity
has no diverging influence on physical observers that move along
timelike trajectories.

It is well known that eternal charged and spinning black holes are
characterized by timelike singularities and inner Cauchy horizons
\cite{Chan} which, according to the weak version of the cosmic
censorship conjecture \cite{HawPen,Pen}, are covered by larger event
horizons. Interestingly, the presence of Cauchy horizons inside the
canonical black-hole solutions of the Einstein field equations
signals a potential breakdown of determinism within classical
general relativity. In particular, observers crossing a Cauchy
horizon on a timelike trajectory enter into a spacetime region
inside the black hole in which past directed null geodesics may
terminate on the inner singularity \cite{M1,M2}. The future history
of these observers cannot be determined uniquely from the initial
data and the classical Einstein field equations. This physically
intriguing fact implies that eternal charged and rotating black-hole
spacetimes with inner Cauchy horizons are not globally hyperbolic
\cite{Chan,M1,M2}.

Nevertheless, it has been proved that the mass-inflation mechanism
\cite{M1,M2,M3,M4,M5,M6,M7,M8}, associated with the exponential
blue-shift amplification of matter and radiation fields inside black
holes, turns the pathological Cauchy horizons of asymptotically flat
black holes into spacetime singularities, inner boundaries of the
spacetimes beyond which the future evolution governed by the
classical field equations ceases to make sense. The mass-inflation
scenario \cite{M1,M2,M3,M4,M5,M6,M7,M8} is based on the fact that
physically realistic black-hole spacetimes, which are formed
dynamically from the collapse of self-gravitating matter
configurations with generic initial conditions, are characterized by
the unavoidable presence of time-dependent remnant fields which, in
the exterior spacetime regions, decay asymptotically as an inverse
power of time \cite{Price,HodPir,Notepl}:
\begin{equation}\label{Eq1}
\psi(t\to\infty)\sim t^{-p}\  .
\end{equation}
The associated energy flux of the infalling fields along the black-hole event horizon decays
accordingly as an inverse power of the standard advanced null
coordinate $v$ \cite{M1,Notepd}: ${\cal F}(v\to\infty)\sim
v^{-2(p+1)}$.

These external remnant fields are then infinitely blue-shifted as
they fall into the newly born charged and rotating black holes and
propagate parallel to their inner Cauchy horizons. This blue-shift
mechanism \cite{M1,M2,M3,M4,M5,M6,M7,M8} turns the Cauchy horizons
into singular hypersurfaces. In particular, the radiation flux of the infalling remnant
fields, as measured by observers crossing the Cauchy horizon, is
characterized by the exponentially divergent functional relation
\cite{M1}
\begin{equation}\label{Eq2}
{\cal F}\sim v^{-2(p+1)}\times e^{2\kappa_- v}\to\infty\ \ \ \
\text{for}\ \ \ \ v\to\infty\  ,
\end{equation}
where $\kappa_-$ is the surface gravity which characterizes the
inner Cauchy horizon of the black-hole spacetime \cite{Notekm}.

Interestingly, fully non-linear numerical simulations of the
collapse of self-gravitating charged scalar fields in asymptotically
flat spacetimes \cite{M6,M7,M8} have explicitly demonstrated that
the inner Cauchy horizons of dynamically formed black holes are
transformed into weak null singularities which are connected to the
strong spacelike singularities of the central $r=0$ hypersurfaces.
Thus, as opposed to eternal black holes, dynamically formed black
holes embedded in asymptotically flat spacetimes contain no
pathological timelike singularities. These physically realistic
black-hole spacetimes therefore respect the Penrose strong cosmic
censorship conjecture \cite{HawPen,Pen}.

Intriguingly, it has recently been claimed \cite{Car1} that {\it
non}-asymptotically flat charged Reissner-Nordstr\"om-de Sitter
(RNdS) black holes may provide a genuine counter-example to the
Penrose SCC conjecture. The physically interesting assertion made in
\cite{Car1} is based on the fact that, as opposed to asymptotically
flat dynamical black-hole spacetimes which are characterized by
external inverse {\it power-law} decaying tails [see Eq.
(\ref{Eq1})], non-asymptotically flat spacetimes are characterized
by {\it exponentially} decaying remnant fields \cite{M1,Car1,Btt}.
In particular, it has been demonstrated \cite{M1,Car1,Btt} that the
relaxation phase of neutral perturbation fields in the exterior
spacetime regions of charged RNdS black holes is dominated by
exponentially decaying tails of the form
\begin{equation}\label{Eq3}
\psi(t\to\infty)\sim e^{-g t}\  ,
\end{equation}
where the spectral gap parameter $g=g(M,Q,\Lambda)$ depends on the
physical parameters (mass, electric charge, cosmological constant)
of the black-hole spacetime.

The effectiveness of the mass-inflation mechanism
\cite{M1,M2,M3,M4,M5,M6,M7,M8} in transforming the inner Cauchy
horizons of non-asymptotically flat black-hole spacetimes, which are
pathological from the point of view of the SCC conjecture, into
singular non-extendable hypersurfaces depends on a delicate
competition between two opposite physical mechanisms: (1) the decay
of remnant perturbation fields in the exterior regions of the
dynamically formed black-hole spacetimes, and (2) the exponential
blue-shift amplification of the infalling fields inside the black
holes. These two physical mechanisms are respectively controlled by
the physical parameters $g$ and $\kappa_-$ [see Eqs. (\ref{Eq2}) and
(\ref{Eq3})].

Specifically, the fate of the inner Cauchy horizons inside
dynamically formed non-asymptotically flat black-hole spacetimes
depends on the dimensionless physical ratio
\cite{Car1,CHe1,CHe2,CHe3}
\begin{equation}\label{Eq4}
\Gamma\equiv {{g}\over{\kappa_-}}\  .
\end{equation}
In particular, as explicitly shown in \cite{Car1,CHe1,CHe2,CHe3}, if
there exists a finite range of the black-hole physical parameters
$\{M,Q,\Lambda\}$ for which
\begin{equation}\label{Eq5}
\Gamma(M,Q,\Lambda)>{1\over2}\  ,
\end{equation}
then the corresponding black-hole spacetimes are physically
extendable beyond their inner Cauchy horizons, a pathological fact
which signals a breakdown of determinism (and therefore a violation
of the fundamental Penrose SCC conjecture) within classical general
relativity.

The physically interesting numerical study of Cardoso, Costa,
Destounis, Hintz, and Jansen \cite{Car1} has recently revealed the
intriguing fact that the decay of {\it neutral} perturbation fields
in near-extremal (highly charged) RNdS black-hole spacetimes is
characterized by the pathological dimensionless relation
\begin{equation}\label{Eq6}
{1\over2}<\Gamma(M,Q,\Lambda)\leq1\ \ \ \ \text{for {\it neutral}
massless fields}\  .
\end{equation}
It was therefore asserted in \cite{Car1,PRLV} that
non-asymptotically flat near-extremal charged RNdS black-hole
spacetimes may provide physically interesting counter-examples to
the fundamental Penrose SCC conjecture.

The main goal of the present paper is to reveal the physically
important fact that, taking into account the unavoidable presence of
{\it charged} fields in dynamically formed {\it charged} black-hole
spacetimes, the seemingly pathological RNdS spacetimes actually
respect the Penrose SCC conjecture \cite{HawPen,Pen}. To this end,
we shall study below the linearized relaxation dynamics of newly
born charged RNdS black holes. In particular, we shall use
analytical techniques in order to calculate the characteristic
quasinormal resonant frequencies (the characteristic damped
oscillations) of linearized charged massive scalar fields in the
charged RNdS black-hole spacetime. Interestingly, below we shall
explicitly prove that, as opposed to the {\it neutral} perturbation
fields considered in \cite{Car1}, the relaxation phase of {\it
charged} fields in the large-coupling regime $qQ\gg1$ \cite{Noteqqb}
is characterized by the physically acceptable dimensionless relation
$\Gamma(M,Q,\Lambda)<1/2$ [see Eq. (\ref{Eq31}) below].


{\it Description of the system.---} We analyze the linearized
dynamics of charged massive scalar fields in the non-asymptotically
flat charged Reissner-Nordstr\"om-de Sitter spacetime, whose
spherically symmetric curved line element is given by
\cite{Chan,Kono1,Noteunit}
\begin{equation}\label{Eq7}
ds^2=-f(r)dt^2+{1\over{f(r)}}dr^2+r^2(d\theta^2+\sin^2\theta
d\phi^2)\ \ \ \ \ \text{with}\ \ \ \ \
f(r)=1-{{2M}\over{r}}+{{Q^2}\over{r^2}}-{{\Lambda r^2}\over{3}}\  .
\end{equation}
Here $\{M,Q\}$ are respectively the mass and electric charge of the
central black hole \cite{NoteQ}, and $\Lambda>0$ is the cosmological
constant of the spacetime. The zeroes of the radial metric function,
\begin{equation}\label{Eq8}
f(r_*)=0\ \ \ \ \text{with}\ \ \ \ *\in\{-,+,\text{c}\}\  ,
\end{equation}
determine the radii $\{r_-,r_+,r_{\text{c}}\}$ of the inner
(Cauchy), outer (event), and cosmological horizons which
characterize the non-asymptotically flat charged black-hole
spacetime \cite{Noterr}.

The dynamics of charged massive scalar fields in the curved RNdS
spacetime is determined by the Klein-Gordon differential equation
\cite{HodPir,Kono1,Stro}
\begin{equation}\label{Eq9}
[(\nabla^\nu-iqA^\nu)(\nabla_{\nu}-iqA_{\nu})-\mu^2]\Psi=0\  ,
\end{equation}
where $\{\mu,q\}$ are respectively the proper mass and the charge
coupling constant of the field
, and
$A_{\nu}=-\delta_{\nu}^{0}{Q/r}$ is the electromagnetic potential of
the charged black-hole spacetime. Expanding the charged massive
scalar field $\Psi$ in the form
\begin{equation}\label{Eq10}
\Psi(t,r,\theta,\phi)=\int\sum_{lm}{{\psi_{lm}(r;\omega)}\over{r}}Y_{lm}(\theta)e^{im\phi}e^{-i\omega
t} d\omega\ ,
\end{equation}
and using the differential relation $dy={{dr}/{f(r)}}$
for the tortoise coordinate $y$, one finds from Eqs. (\ref{Eq7}) and
(\ref{Eq9}) that the radial scalar function is determined by the
Schr\"odinger-like ordinary differential equation \cite{Noteom}
\begin{equation}\label{Eq11}
{{d^2\psi}\over{dy^2}}+V\psi=0\  .
\end{equation}
The effective radial potential, $V=V(r;M,Q,\Lambda,\omega,q,\mu,l)$,
which characterizes the composed
charged-RNdS-black-hole-charged-massive-scalar-field system, is
given by the compact functional expression \cite{Kono1}
\begin{equation}\label{Eq12}
V(r)=\Big(\omega-{{qQ}\over{r}}\Big)^2-{{f(r)G(r)}\over{r^2}}\ \ \ \
\ \text{with}\ \ \ \ \
G(r)=\mu^2r^2+l(l+1)+{{2M}\over{r}}-{{2Q^2}\over{r^2}}-{{2\Lambda
r^2}\over{3}}\ .
\end{equation}

The quasinormal resonant frequencies of the charged scalar fields in
the non-asymptotically flat charged black-hole spacetime (\ref{Eq7})
are characterized by purely ingoing waves at the black-hole outer
horizon and purely outgoing waves at the cosmological horizon
\cite{Kono1}:
\begin{equation}\label{Eq13}
\psi \sim
\begin{cases}
e^{-i(\omega-qQ/r_+)y} & \text{\ for\ \ \ } r\rightarrow r_+\ \
(y\rightarrow -\infty)\ ; \\ e^{i(\omega-qQ/r_{\text{c}})y} & \text{
for\ \ \ } r\rightarrow r_{\text{c}}\ \ \ (y\rightarrow \infty)\  .
\end{cases}
\end{equation}
These physically motivated boundary conditions for the  eigenmodes
of the charged scalar fields determine the discrete spectrum
$\{\omega_n(M,Q,\Lambda,q,\mu,l)\}_{n=0}^{n=\infty}$ of complex
quasinormal resonant frequencies which characterize the composed
RNdS-black-hole-charged-massive-scalar-field system.


{\it The quasinormal resonance spectrum of the composed
charged-RN$d$S-black-hole-charged-field system.---}In the present
section we shall determine the discrete set of complex (damped)
resonant frequencies which characterize the linearized dynamics of
the charged massive scalar fields in the non-asymptotically flat
charged RNdS black-hole spacetime. In particular, as we shall
explicitly show below, the resonant spectrum can be determined {\it
analytically} in the dimensionless large-coupling regime
\cite{Noteqqb,Noteqqmr}
\begin{equation}\label{Eq14}
\alpha\equiv qQ\gg\text{max}\{\mu r_+, l+1\}\
\end{equation}
of the composed charged-black-hole-charged-scalar-field system. In
addition, as explicitly shown in \cite{Schw1,Schw2,Schw3,Schw4}, the
Schwinger-type pair-production mechanism of charged particles in the
charged black-hole spacetime yields the upper bound $Q/r^2_+\ll
\mu^2/q\hbar$ on the black-hole electric field, or equivalently
\begin{equation}\label{Eq15}
\alpha\ll\mu^2r^2_+\  .
\end{equation}

In the dimensionless regime (\ref{Eq14}), the radial potential
(\ref{Eq12}), which determines the dynamics of the charged massive
scalar fields in the charged RNdS black-hole spacetime, has the form
of an effective potential barrier whose fundamental quasinormal
resonant modes can be determined analytically using standard WKB
techniques \cite{WKB1,WKB2,WKB3}. In particular, as we shall now
show, in the large-coupling regime (\ref{Eq14}), the peak $r=r_0$ of
the effective potential barrier (\ref{Eq12}) is located in the
near-horizon region
\begin{equation}\label{Eq16}
{{r_0-r_+}\over{r_+-r_-}}\ll1\
\end{equation}
of the charged black-hole spacetime. Defining the dimensionless
physical variables
\begin{equation}\label{Eq17}
x\equiv{{r-r_+}\over{r}}\ \ \ \ ; \ \ \ \
\varpi\equiv\omega\cdot{{r_+}\over{\alpha}}-1\ ,
\end{equation}
one can express the effective black-hole-charged-scalar-field
potential (\ref{Eq12}) in the compact dimensionless form
\begin{equation}\label{Eq18}
r^2_+V(x;\varpi)=[\alpha(x+\varpi)]^2-2G_0\beta\cdot x [1+O(x/\beta)]\ ,
\end{equation}
where $G_0\equiv G(x=0)$ \cite{Noteie1}, and the dimensionless
parameter $\beta$ is defined by the gradient relation \cite{Notekp}
\begin{equation}\label{Eq19}
\beta\equiv {1\over2}{{df(x=0)}\over{dx}}= \kappa_+ r_+\  .
\end{equation}
From Eq. (\ref{Eq18}) one finds that the peak location, $x=x_0$, of
the composed charged-black-hole-charged-scalar-field scattering
potential is characterized by the simple dimensionless relation
\cite{Noteleq}
\begin{equation}\label{Eq20}
x_0+\varpi={{G_0\beta}\over{\alpha^2}}\ll 1\  .
\end{equation}

It has been explicitly proved in \cite{WKB1,WKB2} that the
characteristic quasinormal resonant frequencies of the
Schr\"odinger-like radial differential equation (\ref{Eq11}) are
determined by the WKB resonance equation
\begin{equation}\label{Eq21}
iK=n+{1\over 2}+\Lambda(n)+O[\Omega(n)]\ \ \ \ ; \ \ \ \ n=0,1,2,...
\end{equation}
where \cite{WKB2}
\begin{equation}\label{Eq22}
K={{V_0}\over{\sqrt{2V^{(2)}_0}}}\ \ \ \ ; \ \ \ \
\Lambda(n)={{1}\over{\sqrt{2V^{(2)}_0}}}\Big[{{1+(2n+1)^2}\over{32}}\cdot{{V^{(4)}_0}\over{V^{(2)}_0}}-
{{28+60(2n+1)^2}\over{1152}}\cdot\Big({{V^{(3)}_0}\over{V^{(2)}_0}}\Big)^2\Big]\
,
\end{equation}
and the cumbersome mathematical expression of the sub-leading
correction term $\Omega(n)$ is given by equation (1.5b) of
\cite{WKB2}. Here the spatial derivatives $V^{(k)}_0\equiv
d^{k}V/dy^{k}$ of the effective black-hole-field potential $V(y)$
are evaluated at the radial location $y=y_0$ of its scattering peak.

Substituting Eqs. (\ref{Eq18}) and (\ref{Eq20}) into Eq.
(\ref{Eq22}), one finds the functional expressions \cite{Noteie1}
\begin{equation}\label{Eq23}
K={{G_0}\over{2\alpha}}\cdot{{\varpi-{{G_0\beta}\over{2\alpha^2}}}
\over{ {{G_0\beta}\over{\alpha^2}}-\varpi}}\ \ \ \ ; \ \ \ \
\Lambda(n)=-{{2\alpha}\over{G_0}}\cdot (n+{1\over 2})^2\ \ \ \ ; \ \
\ \ \Omega(n)=O[(\alpha/G_0)^2(n+1/2)^3]\
\end{equation}
for the various terms in the WKB resonance equation (\ref{Eq21}).
From Eqs. (\ref{Eq21}) and (\ref{Eq23}), one obtains the (rather
cumbersome) expressions
\begin{equation}\label{Eq24}
\varpi_R={{G_0\beta}\over{2\alpha^2}}\cdot
{{G^2_0+2(2n+1)^2\alpha^2[1-(2n+1)\alpha/G_0]^2}\over{G^2_0+(2n+1)^2\alpha^2[1-(2n+1)\alpha/G_0]^2}}\cdot
\big\{1+O[(\alpha/G_0)^2]\big\}
\end{equation}
and
\begin{equation}\label{Eq25}
\varpi_I=-i{{G_0\beta}\over{2\alpha^2}}\cdot
{{(2n+1)\alpha G_0[1-(2n+1)\alpha/G_0]}\over{G^2_0+(2n+1)^2\alpha^2[1-(2n+1)\alpha/G_0]^2}}\cdot
\big\{1+O[(\alpha/G_0)^2]\big\}\  .
\end{equation}
for the dimensionless resonant frequencies which characterize the
composed charged-black-hole-charged-scalar-field system in the
large-coupling regime (\ref{Eq14}). Taking cognizance of the strong
inequality [see Eq. (\ref{Eq15}) and \cite{Noteie1}]
\begin{equation}\label{Eq26}
{{\alpha}\over{G_0}}\ll1\  ,
\end{equation}
one obtains from (\ref{Eq24}) and (\ref{Eq25}) the approximated
compact expressions
\begin{equation}\label{Eq27}
\varpi_R={{G_0\beta}\over{2\alpha^2}}\cdot \big\{1+O[(\alpha/G_0)^2]\big\}
\end{equation}
and
\begin{equation}\label{Eq28}
\varpi_I=-i{{\beta}\over{\alpha}}(n+1/2)[1-(2n+1)\alpha/G_0]\cdot
\big\{1+O[(\alpha/G_0)^2]\big\}\ .
\end{equation}

Finally, substituting Eqs. (\ref{Eq27}) and (\ref{Eq28}) into the
relation $\omega=(qQ/r_+)\cdot(\varpi+1)$ [see Eqs. (\ref{Eq14}) and
(\ref{Eq17})], and using the identity (\ref{Eq19}), we obtain the
remarkably compact functional expression
\begin{equation}\label{Eq29}
\omega_n={{qQ}\over{r_+}}+{{G_0\kappa_+}\over{2qQ}}-i\kappa_+\cdot
\Big[n+{1\over
2}-{{2qQ}\over{G_0}}\cdot\Big(n+{{1}\over{2}}\Big)^2\Big]\ \ \ ; \ \
\ n=0,1,2,...
\end{equation}
for the quasinormal resonant spectrum of the composed
charged-RNdS-black-hole-charged-massive-scalar-field system in the
large-coupling regime (\ref{Eq14}).

In particular, the fundamental (least damped) resonant mode of the
system, which dominates the relaxation phase of the dynamically
formed non-asymptotically flat charged black-hole spacetimes, is
characterized by the simple relation
\begin{equation}\label{Eq30}
\Im\omega_0={1\over2}\kappa_+\Big(1-{{qQ}\over{G_0}}\Big)\ .
\end{equation}


{\it Summary and discussion.---}The Penrose strong cosmic censorship
conjecture \cite{HawPen,Pen} has attracted the attention of
physicists and mathematicians during the last five decades. This
physically intriguing conjecture asserts that, starting with generic
initial conditions, the future evolution of the non-linearly coupled
matter-curvature fields should always produce globally hyperbolic
spacetimes.


As nicely shown in \cite{Car1,CHe1,CHe2,CHe3}, the (in)stability
properties of the inner Cauchy horizons of dynamically formed
charged RNdS black holes, which ultimately determine the fate of the
Penrose SCC conjecture in these non-asymptotically flat spacetimes,
are controlled by the simple dimensionless ratio
$\Gamma(M,Q,\Lambda)\equiv g/\kappa_-$ [see Eqs. (\ref{Eq3}) and
(\ref{Eq4})] between the decay rate $g$ of remnant perturbation
fields in the exterior regions of the black-hole spacetime and the
blue-shift growth rate $\kappa_-$ of the infalling fields inside the
black holes.

In particular, as explicitly demonstrated in
\cite{Car1,CHe1,CHe2,CHe3}, non-asymptotically flat black-hole
spacetimes which are characterized by the inequality $\Gamma>1/2$
are physically extendable beyond their inner Cauchy horizons, a fact
which signals the breakdown of determinism and the corresponding
violation of the Penrose SCC conjecture in these classical
black-hole spacetimes.

Intriguingly, the recently published important numerical study of
Cardoso, Costa, Destounis, Hintz, and Jansen \cite{Car1} has
revealed the fact that the decay rate of {\it neutral} perturbation
fields in the exterior spacetime regions of highly charged
(near-extremal) RNdS black holes is too fast to make their inner
Cauchy horizons unstable. In particular, it has been explicitly
proved that there is a finite volume in the phase space of the
physical parameters $\{M,Q,\Lambda,l,m\}$, which characterize the
composed charged-black-hole-neutral-scalar-field system, for which
the dimensionless ratio $g/\kappa_-$ is characterized by the
pathological relation $1/2<\Gamma(M,Q,\Lambda)\leq1$ \cite{Car1}. It was
therefore concluded in \cite{Car1} that these non-asymptotically
flat black-hole spacetimes may provide physically interesting
counter-examples to the fundamental SCC conjecture.

It should be realized that the existence of even one physically
acceptable counter-example to the Penrose SCC conjecture would imply
the undesirable breakdown of determinism in Einstein's classical
general theory of relativity. It is therefore physically crucial to
restore the predictive power of the Einstein field equations by
proving that, due to the existence of some physical mechanism that
has not been taken into account in previous analyzes, the
dynamically formed RNdS black-hole spacetimes, which have been
suspected to violate the Penrose SCC conjecture, are globally
hyperbolic.

In the present paper we have pointed out that the presence of
charged remnant fields in the exterior spacetime regions of
dynamically formed RNdS black holes is an unavoidable feature of
these non-asymptotically flat charged spacetimes. In particular, we
have demonstrated that, as opposed to the {\it neutral} remnant
fields considered in \cite{Car1}, the decay rates of {\it charged}
remnant fields \cite{Notehs,HSF} in these dynamically formed
non-asymptotically flat charged black-hole spacetimes are slow
enough to guarantee, through the mass-inflation mechanism
\cite{M1,M2,M3,M4,M5,M6,M7,M8}, the instability of the inner Cauchy
horizons. Specifically, using analytical techniques, it has been
explicitly proved that the relaxation phase of the composed
charged-black-hole-{\it charged}-scalar-field system in the
large-coupling regime $qQ\gg1$ \cite{Noteqqb} is characterized by
the physically acceptable dimensionless relation [see Eq.
(\ref{Eq30})] \cite{Notekps}
\begin{equation}\label{Eq31}
\Gamma(M,Q,\Lambda)<{{\kappa_+}\over{2\kappa_-}}\leq{1\over2}\ \ \ \
\text{for {\it charged} massive fields}\  .
\end{equation}

The analytically derived results presented in this paper therefore
reveal the physically important fact that, taking into account the
unavoidable presence of charged remnant fields in dynamically formed
charged RNdS black holes, these non-asymptotically flat spacetimes
respect the Penrose strong cosmic censorship conjecture
\cite{HawPen,Pen}.

\bigskip
\noindent
{\bf ACKNOWLEDGMENTS}
\bigskip

This research is supported by the Carmel Science Foundation. I would
like to thank Yael Oren, Arbel M. Ongo, Ayelet B. Lata, and Alona B.
Tea for helpful discussions.



\begin{thebibliography}{99}
\bibitem{Notepl} The power-law index $p$, whose precise value is not important for our analysis,
depends on the angular harmonic index $l$ of the field mode and, for
charged fields, also on the dimensionless charge coupling parameter
$\alpha\equiv qQ$ (here $Q$ and $q$ are respectively the electric
charge of the black hole and the charge coupling constant of the
fields) \cite{Price,HodPir}.
\bibitem{Notepd} Note that the energy flux of the radiation fields
is proportional to $(\dot \psi)^2$ \cite{M1}.
\bibitem{Notekm} The surface gravities, which characterize the horizons
of the black-hole spacetime, are directly related to the gradients
of the radially dependent metric function $g_{00}$ [see Eq.
(\ref{Eq19}) below].
\bibitem{Noteqqb} It is worth noting that in our universe with $q^2/\hbar\simeq 1/137$
the relation $qQ\gg1$ is satisfied by even slightly charged black
holes which contain a tiny number $N\gtrsim 10^3$ of elementary
charged particles.
\bibitem{Noteunit} We shall henceforth use natural units in which $G=c=\hbar=1$.
\bibitem{NoteQ} We shall assume $Q>0$ without loss of generality.
\bibitem{Noterr} Note that the horizons of the black-hole spacetime
are characterized by the relations $r_-\leq r_+\leq r_{\text{c}}$.
\bibitem{Noteom} For brevity, we shall henceforth omit the harmonic indices $\{l,m\}$
of the resonant eigenmodes which characterize the charged massive
scalar fields in the charged black-hole spacetime.
\bibitem{Noteqqmr} Note that charged elementary particles in
our universe are characterized by the strong dimensionless
inequality $\mu/q\ll1$. Thus, the relation $qQ\gg \mu r_+$ is
satisfied by even slightly charged black holes with $Q/r_+ \ll1$.
\bibitem{Noteie1} Note that the strong inequalities (\ref{Eq14}) and (\ref{Eq15}) imply
the dimensionless relation $\mu r_+\gg1$, and thus $G_0\simeq \mu^2 r^2_+[1+O(1/\mu^2r^2_+)]$.
\bibitem{Notekp} Here $\kappa_+$ is the surface gravity which
characterizes the event horizon of the charged black-hole spacetime.
\bibitem{Notehs} It is worth pointing out that spin-dependent
effects become negligible in the large-coupling regime (\ref{Eq14})
\cite{HSF}. Thus, our analytically derived results are expected to
be valid for charged higher-spin fields as well.
\bibitem{Notekps} Here we have used the characteristic dimensionless relation
$\kappa_-/\kappa_+\geq1$ \cite{M2} for the surface gravities
$\kappa_i\equiv |df(r=r_i)/dr|/2$ of the black-hole (Cauchy and event) horizons.
\end{thebibliography}
\end{document}